\documentclass[conference]{IEEEtran}
\IEEEoverridecommandlockouts
\usepackage{amsmath,amssymb,amsfonts}
\usepackage{subcaption,color,algorithmic, algorithm}
\usepackage{graphicx}
\usepackage{textcomp}
\usepackage{xcolor}
\def\BibTeX{{\rm B\kern-.05em{\sc i\kern-.025em b}\kern-.08em
    T\kern-.1667em\lower.7ex\hbox{E}\kern-.125emX}}
\begin{document}

\title{\LARGE \bf Survivable Networks via UAV Swarms Guided by Decentralized Real-Time Evolutionary Computation}

%



\author{\IEEEauthorblockN{George Leu}
\IEEEauthorblockA{\textit{School of Engineering and Information Technology} \\
\textit{UNSW Canberra}\\
Canberra, Australia \\
g.leu@adfa.edu.au}
\and
\IEEEauthorblockN{Jiangjun Tang}
\IEEEauthorblockA{\textit{School of Engineering and Information Technology} \\
\textit{UNSW Canberra}\\
Canberra, Australia \\
j.tang@adfa.edu.au}
}

\maketitle

\begin{abstract}
    The \textit{survivable network} concept refers to contexts where the wireless communication between ground agents needs to be maintained as much as possible at all times, regardless of any adverse conditions that may arise. 
    In this paper we propose a nature-inspired approach to survivable networks, in which we bring together swarm intelligence and evolutionary computation. We use an on-line real-time Genetic Algorithm to optimize the movements of an UAV swarm towards maintaining communication between the ground agents. The proposed approach models the ground agents and the UAVs as boids-based swarms, and optimizes the movement of the UAVs using different instances of the GA running independently on each UAV. The UAV coordination mechanism is an implicit one, embedded in the fitness function of the Genetic Algorithm instances. The behaviors of the individual UAVs emerge into an aggregated optimization of the overall network survivability. The results show that the proposed approach is able to maintain satisfactory network survivability levels regardless of the ground agents' movements, including for cases as complex as random walks. 
    
\end{abstract}

\section{Introduction}

In survivable networks domain there is currently a growing interest to employ swarms of UAVs that act as relays for maintaining wireless communication between the agents of a ground swarm. Network survivability refers to the extent to which the group of ground agents remains connected or recovers from connectivity loss, when obstructions for the wireless communication occur~\cite{Hunjet2017,Hui2017}. These obstructions can be actual physical obstacles (i.e. mountains, forests, buildings), or, simply, the distance between the ground agents, which naturally attenuates the radio signals. The use of UAVs, especially rotary wing UAVs, as relays for the ground agents is gaining great popularity at both theoretical and practical levels, in application fields like search and rescue in disaster areas, large scale farming operations, military operations, internet of things, autonomous cars and transportation, and many others. In such contexts, the UAVs continuously adjust their positions to the movement pattern of the ground agents, so that the ground communication is maintained as much as possible at all times. The main challenges associated to the above are related to the mobility models employed for implementing the movement of the individual UAVs, and to the coordination mechanisms available for the individual UAVs to perform well collectively as a relay swarm~\cite{ZengUAVCom2016}.

In this paper, we use a nature-inspired approach that combines swarm intelligence and evolutionary computation to provide ground network survivability regardless of the ground agents' movements. To address the challenges mentioned above, we propose an implementation of the UAVs' mobility which is inspired from the classic boids model of Reynolds~\cite{reynolds1987flocks}. The proposed model uses modified versions of the two key concepts of the classic boids: the interaction based on the concept of neighborhood, and the position update rule based on the weighted sum of a set of primitive forces. The modifications we propose are that the neighborhood is network-based~\cite{Tang2018} instead of the classic vision-based~\cite{reynolds1987flocks}, and the update rule includes forces from outside the swarm, not only from within. The influence from outside comes from the ground agents, whose movements have to be tracked by the UAVs. Further, to couple the proposed mobility model to the survivable networks problem, we propose an on-line real-time genetic algorithm that optimizes the weights in the update rule towards maximizing the network survivability. The genetic algorithm runs in multiple independent instances in parallel (i.e. one instance on each UAV), and embeds in its fitness function a network coverage metric. The way the fitness function is defined contributes to addressing the coordination challenge too. By attempting to maximize their individual coverages, the UAVs take individual quasi-greedy actions which aggregate to form an implicit coordination mechanism that maximizes the ground network connectivity (which is the metric we use to evaluate the network survivability).

In addition to the main contributions mentioned above, we also contribute to the mobility model of the ground agents, which we implement using Reynold's classic boids, with no modification of the key concepts. This leads to a \textit{dual boids-based swarm with evolution} view on survivable networks, where the ground swarm operates independently to simulate a certain mission in the field, and the airborne swarm operates as a networked boids-based swarm with inputs from both air and ground agents. This view offers two other benefits, which are typically not found in the evolutionary swarm applications related to survivable networks. The first benefit of considering the dual swarm is that the number of parameters used to model the system is overall very low, facilitating evolutionary optimization approaches in general, and our proposed GA in particular, to run in real-time. Thus, very complex non-deterministic behaviors can be obtained by optimizing only a very small number of weights, associated with the boids forces that implement the swarm behavior. The second benefit, which flows from the first one, is the scalability of the system. Due to the small number of primitive forces typically involved by boids-based swarms, the number of agents used in experiments (both airborne and ground) is virtually irrelevant. Thus, very large systems, with very complex behaviors can be investigated with fairly low computational power required.

With the proposed approach, we perform experiments using the swarm of ground agents in two contexts. First, the ground agents operate as pure classic boids, and then as random walkers. The former case implements a very general movement pattern, while the latter implements no movement pattern at all. Arguably, the less pattern exists in the movement of the ground agents, the more difficult it is for the UAVs to track them and provide network survivability. Thus, ideally, it would be desired that the UAVs provide ground network survivability regardless of the ground agents' movements (i.e. with or without a coherent pattern). While that is not the case in practice, we demonstrate in our experiments that the proposed real-time GA, in conjunction with the boids-based UAV mobility model, is able to provide very good results for the boids-based ground movement and also satisfactory results for the random walks. In summary, the results we obtain offer convincing evidence that the methods employed ensure network survivability for complex ground activity, including good responsiveness to activities that have no pattern at all, such as random walks.
 

\section{BACKGROUND} \label{Section:Background}

Historically, the network survivability problems have been first approached using ground relays, either fixed or mobile. Later, high-altitude airborne relays like satellites and fixed-wing aircraft have been used. However, both the ground and the high-altitude approaches have substantial limitations in numerous respects. The very first issue is their limited ability to reach (or follow) the ground agents in difficult locations, such as densely forested areas, densely built urban areas, indoors, or underground; that is, a significant mobility issue. Other limitations, which are equally important, are the difficulties in scaling the systems, and also the technological complexity of the hardware and communication technologies employed. This has led to studies that involved very low number of agents, where the very concept of swarms is not really applicable. Typical scenarios in these studies implement very low scale systems with 1 relay agent and 2 ground agents~\cite{Elliot2016,Hui2017}, few relay agents to support communication between a single mobile ground agent and a fixed base station~\cite{Hauert2009}, or few relay agents and few ground agents~\cite{BasuUAV2004,Hui2011,ZhouAGCop2015,Hunjet2017}.

More recently, the advances in drone technology allowed the use in these contexts of miniaturized rotary-wing UAVs (i.e. multicopter drones), which provide high mobility and versatility at relatively low cost. This allowed for a wider range of mobility models to be employed for the UAVs, which in turn facilitated the investigation of a wider range of air-ground systems. The mobility models employed by the UAVs refer to the individual control and the group-level cooperation mechanisms that allow the UAVs to find, either individually or collectively, the optimal air trajectories/positions for maximizing the communication capabilities of the ground agents.

Early studies on mobility discussed models as simple as random mobility~\cite{Kuiper2006}, which later generated the more complex concept of chaos enhanced mobility ~\cite{Rosalie2016,Rosalie2017,Rosalie2018}. Other studies proposed analytic and/or parametric approaches with fixed pre-tuned parameters ~\cite{KarInria2003,BasuUAV2004,Hui2011,Cetin2012,Goddemeier2012,ZhouAGCop2015,Zhang2018}. However, both these directions of research suffered from limitations related to scalability. 

More recently, various nature-inspired algorithms have been used in the mobility models, such as ant colony pheromone-based mobility~\cite{Kuiper2006}, bat algorithms~\cite{SUAREZ2018}, or boids-based finite state machines~\cite{BasuUAV2004}. Evolutionary computation, which is of particular interest for this paper, has been also employed in various forms in UAV path planning and coordination. One class of studies used evolution as standalone method for the mobility, where evolutionary algorithms evolve parameters of the controllers used in the individual UAVs~\cite{Hauert2009}. These studies do not account for the collective behavior of the UAVs, that is, they apply only to individual agents, with no coordination mechanisms in place. Another class of studies employs evolutionary algorithms that contribute to both the individual and group mobility of the UAVs~\cite{Gaudiano2005,Lamont2007,Roberge2013}.

One important issue of the evolutionary algorithms in these contexts is their limited ability to operate on-line, in real-time, due to the relatively slow convergence~\cite{Tang2012}. Thus, most of the studies report off-line and/or centralised algorithms in relation to the survivable networks domain. In~\cite{Gaudiano2005} the authors use a centralised Genetic Algorithm to evolve parameters for the behavior of UAVs, where the behavior of the UAVs is based on a small set of primitive individual actions and transitions (5 possible actions and 11 transitions), however, the coordination mechanism requires global knowledge, or at least a persistent trace left in the environment by each UAV. Lamont et. al~\cite{Lamont2007} use a multi-objective evolutionary algorithm that operates quasi locally, based on the pinning concept from control theory. The objectives are related to the cost of operation (distance distance traveled and amount of climbing), and to the risk resulting from flying through difficult areas. The solution set offers paths that provide the lowest cost associated with a particular level of risk.

To alleviate the real-time operation issue, Particle Swarm Optimization (PSO) methods have been proposed, or hybrid methods that combine GAs and PSO. In~\cite{Roberge2013}, a comparison between the two directions is discussed, where the authors show that the speed of PSOs and GAs can vary greatly depending on the type scenario they are employed for, however, in most of the scenarios the PSO is faster, and therefore more appropriate for real-time use. PSO's speed gain over GAs is also reported by Tang et. al~\cite{Tang2012}, who use a PSO method for UAV coordination, and also by Duan et. al~\cite{Duan2013}, who use a hybrid PSO-GA for multi-UAV formation control.

In spite of the issue discussed above, the GA we propose in this paper is fast enough to operate in real-time for the scenarios considered, which include 4 UAVs and 100 ground agents. While the GA itself is designed to operate in real-time, and therefore is fast, there is also the boids-based mobility model we propose, which contributes to its speed via the low number of parameters to be optimized. Another contribution to the speed of the GA, and also to the scalability of the system, is the level of integration we adopt for the UAVs' and ground agents' behaviors. As far as we know, there is no attempt in the current research to implement both the ground agents and the UAVs using the same conceptual model. The existing studies consider the airborne and ground agents as separate groups, and as a result, their operation and mobility are implemented using different conceptual models. By considering the ground agents as a boids-based swarm too, we integrate seamlessly the air-ground operation (i.e. two boids-based swarms with slightly different settings of the parameters). Thus, the parameters in the GA that come as inputs from the ground swarm are only a few (i.e. 2), and of exactly the same type as those corresponding to the UAV swarm. The details of the proposed methodology are provided below, in Section~\ref{Section:EnvSwarmSurv}.

\section{Methdology}\label{Section:EnvSwarmSurv}

In this paper, we consider that the UAVs fly at a constant low altitude that does not affect their communication with the ground. Thus, the vertical distance between the UAVs and ground level can be neglected from a networking perspective, as there is no variation in communication conditions due to altitude changes. For the ground agents, we consider that they operate on a flat field with no obstacles. This means they can move freely across the field, with the only restriction being to avoid collisions with each-other. This type of environment has been successfully used as a test-bed in other studies on survivable networks~\cite{BasuUAV2004,Gaudiano2005}.

We model the swarming behaviours of the two types of agents, airborne and ground, based on the classic boids model of Reynolds~\cite{Reynolds1987}, which means, all agents, airborne and ground, are flocking boids. We use the two key concepts of the original boids model, i.e. the neighborhood-based interaction and the three boids forces: cohesion, alignment, and separation. However, we propose an enhanced version of these concepts, and we enrich these concepts to suit our approach to the survivable networks context.

In the following sections, we describe in detail the methodology used for implementing the ground and UAVs, with the subsequent optimization algorithm, and the metrics used for evaluating the network survivability.

\subsection{Measuring Survivability} \label{Subsection:EvalMetrics}
\begin{figure}[h]
 \centering
    \begin{subfigure}[b]{0.35\textwidth}
        \includegraphics[width=\textwidth]{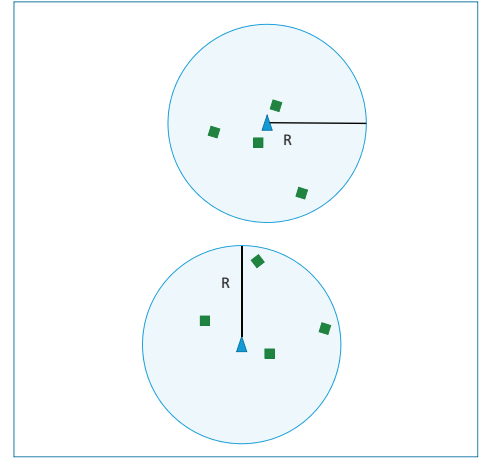}
        \caption{A full coverage network}
        \label{fig:fullconverage}
    \end{subfigure} \\
    \begin{subfigure}[b]{0.35\textwidth}
        \includegraphics[width=\textwidth]{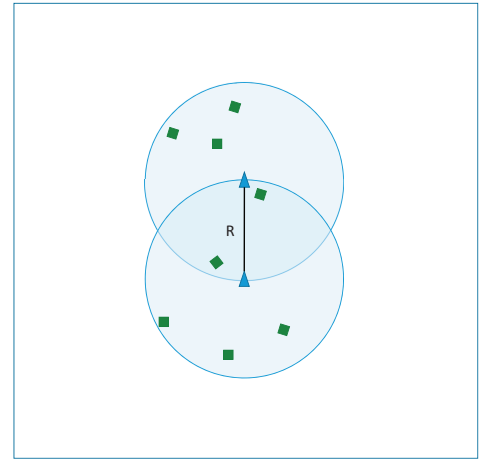}
        \caption{A full connected network}
        \label{fig:fullconnect}
    \end{subfigure}
  \caption{Full coverage v.s. full connectivity, via two UAVs with communication range $R$.}
  \label{Fig:CovCon}
\end{figure}

The ground network survivability can be measured using to major concepts: coverage and connectivity. The coverage refers to the number of ground agents situated within the aggregated area that contains the communication range of all UAVs, as shown in Figure~\ref{fig:fullconverage}. In this case, the ground agents covered by one of the UAVs can communicate between them, but cannot communicate with those covered by the other UAV. As a result, the ground agents do not form a complete connected network; there are two sub-networks where each UAV acts as a relay for its own current local ground network. The coverage is defined very straightforward, as the number of ground agents covered by all UAVs.

The connectivity refers to the situation when the agents are able to communicate regardless of which UAVs' coverage they belong to. Figure~\ref{fig:fullconnect} illustrates a full connectivity case, where a fully connected network is established when all UAVs are close enough to be able to communicate between them, and thus relay the communication between any ground agent. If the ground swarm is fully connected, then it is also fully covered, but not the vice-versa. The connectivity is typically modeled based on the graph theoretical concept of connected network component~\cite{Albert2002}. A connected component is a sub-graph where any two nodes are connected to each other. Therefore, we define the connectivity metric as the number of connected components that exist at a moment in time within the swarm of ground agents. Ideally, the connectivity should have the value $1$, which means there is only one sub-graph which is equal to the entire network. In this case, the ground swarm is fully connected via the UAVs (i.e it operates as an equivalent full connected graph from a graph theory perspective). In practice, less perfect cases are still acceptable, where there are more than one connected components, but among them one or several giant connected components~\cite{Newman2002} exist.

In this paper we use the connectivity as the main metric for evaluating the ground network survivability, and we use the coverage in the fitness function of the optimization algorithm. While this may appear as an inconsistency, we demonstrate that it is possible to use the coverage as the objective for the optimization engine of each individual UAV, and obtain a maximization of the connectivity at the swarm level. This process actually implements an implicit coordination mechanism embedded in the individual goals of the UAVs (this is explained in detail in Section~\ref{Section:Optimization}).

\subsection{Modelling the Ground Agents}

The ground agents implement entirely the classic boids model. They follow the three boids rules, which are applied as a result of the influence received from their ground neighbors, where the neighborhood is defined by agents' vision distance $v_d$ and a vision angle $v_\alpha$. We recall that in survivable networks contexts the ground agents operate in the field to accomplish a certain task, and are not aware or concerned about the existence of the airborne support. Thus, the movement of the ground agents is not influenced by the UAVs. The three boids forces (i.e vectors with magnitude and heading) applied to the ground agents are described below.

\paragraph{Cohesion Force ($C$)} describes the tendency of an agent to move towards its neighbors' location, and is calculated based on the centre of mass (average position) of all agents in its neighborhood.

\paragraph{Alignment Force ($A$)} shows the tendency of the agent to align with the direction of movement of its neighbors, and is calculated based on the average heading of all the neighbors.

\paragraph{Separation Force ($S$)} expresses the tendency of agents to steer away from their neighbours in order to avoid crowding them or colliding with them. The agents need to keep a minimum ground safe distance ($SD_G$) from their neighbors.

Once the forces are calculated based on the neighbours influence, the velocity $V$ of a ground agent at time step $t$ is updated using the following equation:
\begin{equation}
    V(t) = V(t-1) + W_C \cdot C(t) + W_A \cdot A(t) + W_S \cdot S(t)
    \label{Equation:GroundVelocity}
\end{equation}
where $W_C$, $W_A$, and $W_S$ are weights corresponding to the cohesion, alignment, and separation forces. The weights are constant for the ground agents, since their behavior is fixed (i.e. they perform a certain task). Based on the velocity update, the position $P$ of a ground agent at time step $t$ can be updated as follows:
\begin{equation}
    P(t) = P(t-1) + V(t)
    \label{Equation:Position}
\end{equation}

In addition to the boids-based swarming behaviour, we also consider the case when the ground agents move at random. This allows us to investigate the performance of our proposed approach in the most general case, when the ground agents have no movement pattern at all. In the context of boids-based swarming, a random walk movement equates with a swarm in which the cohesion and alignment forces do not exist. However, the agents still need to keep the safe distance from neighbours; hence, the separation force is still applied.

\subsection{Modelling the UAVs}

The UAVs form another boid-based swarm, which embeds the three classic boids forces showing the influence from the neighboring UAVs, and two other additional forces that represent the influence from the neighboring ground agents. Unlike the classic boids model, for the UAVs the neighborhood is not defined by vision, instead we use an omnidirectional communication range $R$. We recall that the purpose in survivable networks is for the UAVs to move according to the ground agents' movement, so that they facilitate communication. This means, they tend to follow/track the ground agents in order to provide the network relay service. Thus, we consider that an UAV is influenced by the movement of the ground agents situated in its neighborhood (i.e. communication range) through the cohesion and alignment forces. The separation force is not applied, since there is never a risk of collision between a ground and an UAV.

The three classic forces applied between UAVs are defined just like the ones for the ground agents, that is, according to the original study of Reynolds~\cite{reynolds1987flocks}. The other two forces, corresponding to the influence from ground agents, are described below. 

\paragraph{Air-Ground Cohesion Force($C_{AG}$)} All ground agents situated within the communication range $R$ of an UAV $A_i$ form the set of ground neighbors $N_G$ of that UAV. Each ground neighbor $g_j \in N_G$ satisfies $dist(A_i, G_j) < R$. Then, the air-ground cohesion force $C_{{AG}_i}$ applied to UAV $A_i$ at time $t$ can be derived from the position of its ground neighbours as in Equation~\ref{Equation:GACohesion}.
\begin{equation}
	C_{{AG}_i} = \frac{\sum_{j=0}^{|N_G|} P_{G_j}}{|N_G|} - P_{A_i} \;\; \mbox{for each} \; G_i \in N_G
	\label{Equation:GACohesion}
\end{equation}
where, $|N_G|$ is the cardinality of $N_G$, $P$ is the position of an agent.

\paragraph{Air-Ground Alignment Force ($A_{AG}$)} The air to ground alignment force ($A_{{AG}_i}$) of an UAV $A_i$ at time $t$ is derived from the velocities of all its ground neighbors $N_G$ as in Equation~\ref{Equation:GAAlignment}. 
\begin{equation}
	A_{{GA}_i} = \frac{\sum_{j=0}^{|N_G|} V_{G_j}}{|N_G|} - V_{A_i} \;\; \mbox{with} \; i\neq j
	\label{Equation:GAAlignment}
\end{equation}
where $|N_G|$ is the cardinality of $N_G$, and $V$ is the velocity of an agent.

Once all five forces are calculated, the velocity $V_{A_i}(t)$ of each UAV $A_i$ is updated as follows:
\begin{equation}
    \begin{split}
        V_{A_i}(t) = & V_{A_i}(t-1) + \\
            & + W_{C_A} C_{A_i}(t) + W_{A_A} A_{A_i}(t) + W_{S_A} S_{A_i}(t)\\
            & + W_{C_{AG}} C_{{AG}_i}(t) + W_{A_{AG}} A_{{AG}_i}(t)
    \end{split}
	\label{Equation:AirVelocity}
\end{equation}
where $W$s denote the weights of the forces in the update rule.

Then, the position $P_{A_i}$ at time $t$ of each UAV $A_i$ can be updated as in Equation~\ref{Equation:AirPosition}. 
\begin{equation}
	P_{A_i}(t) = P_{A_i}(t-1) + V_{A_i}(t)
	\label{Equation:AirPosition}
\end{equation}

The rules considered above for the UAVs allow them to move according to a boids-based swarming behavior, where the behavior is guided by the interaction with both airborne and ground agents. However, the swarm behavior alone does not guarantee optimal connectivity services for the ground agents. Unlike the ground agents, the force weights of the UAVs are not constant, since their movement need to adapt comtinuously to the ground activity. Therefore, an optimization of the force weights at each time step is needed in order to achieve the best connectivity for the ground agents. The optimization algorithm is described in detail in Section~\ref{Section:Optimization}.

\subsection{A Decentralized Real-Time Genetic Algorithm}\label{Section:Optimization}
In this paper, we propose an decentralized real-time genetic algorithm as optimization method for the UAVs' movement. We mentioned above that the force weights need to be optimized. This can be done in two ways. One way is to optimize the weights in the same way for all UAVs. This means that at each time step, the current optimal set of weights is used by all agents. This is similar to employing a centralized optimization for the whole swarm. We are interested to make the UAVs individually adaptive, therefore each UAV runs its own optimization engine to obtain its own set of optimal weights at each time step. Thus, each UAV attempts to optimize independently its own five force weights plus the speed, with the purpose of providing better connectivity to the ground agents. In this case, a coordination mechanism is also required. Our approach is that an explicit coordination mechanism is not needed, instead, the way we define the fitness function of each individual agent leads to an implicit coordination capability.

\begin{figure*}[h]
    \centering
    \includegraphics[width=0.6\textwidth]{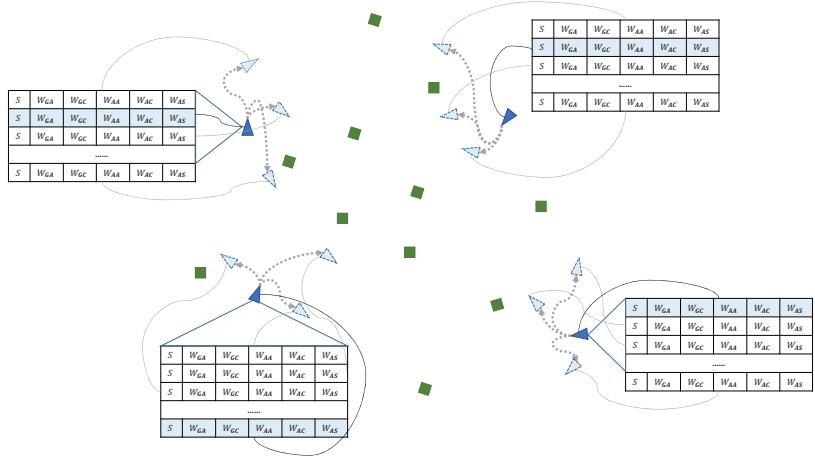}
    \caption{Chromosomes of each UAV in the Decentralized GA. Triangles are UAVs; squares are ground agents}
    \label{fig:DecentralisedEC}
\end{figure*}

Figure~\ref{fig:DecentralisedEC} illustrates how each UAV has its own optimization engine, which relies on a population of chromosomes. At each time step, only one chromosome is active, representing a valid decision in the simulation; all other chromosomes represent shadow agents, which are evolved in the GA but do not take effect in simulation. The structure of the chromosomes is the same for all UAVs, i.e. they are vectors with 6 components ${s,W_{GA}, W_{GC}, W_{AA}, W_{AC}, W_{AS}}$, where $W$s are force weights and $s$ is the speed. The value ranges for the genes in the chromososes are as follows. The speed of an UAV can vary from 0 (hovering) to 5. The separation weight needs to be high in order to maintain safety. Hence, it can take values between 0.5 and 2. All other weights (i.e. $W_{GA}, W_{GC}, W_{AA}, W_{AC}$) can take values from 0 to 0.5.

Ideally, each UAV would run the GA at each time step; however, a certain time is needed for the evolution to reach meaningful results. For this reason the agents run the GA every $t'$ time steps instead of every time step. This means that the GAs run in time windows of duration $t'$, where the duration is the stopping condition. This affects the quality of the optimal solution at each run, but ensures that overall throughout the swarm simulation, the aggregated optimization process runs virtually real-time. The use of GAs in time windows also allows a past/historical period of duration $t'$, as well as a future/prediction period of duration $t'$ to be used for optimization as part of the fitness function. This will be explained below.

Consider an arbitrary agent and an arbitrary time window within the swarm simulation timeline. The agent starts running a GA. In the beginning of the GA, a population of chromosomes is randomly generated within the ranges discussed earlier. One chromosome is selected randomly (illustrated in Figure~\ref{fig:DecentralisedEC} as coloured rows) to be an active chromosome. The active chromosome will be used in the actual swarm simulation, and the speed and force weights will be applied in simulation to produce actual position updates. All other chromosomes represent shadow agents, which update their virtual positions accordingly, but these positions will not reflect in the actual simulation. The shadow chromosomes have their fitness evaluated, a tournament selection with elitism is performed, and one point crossover and mutation are applied for producing the next generation. Then, the best solution among the current population is chosen to become the active agent, and its updates take effect in simulation, while the rest of the chromosomes continue to operate as shadows. The process ends when the time window ends, and the current active chromosome takes effect in the simulation; i.e. it represents the decision of the agent.

The fitness function used in our GA is built upon the number of ground agents covered by an aerial agent and its neighbouring aerial agents (i.e. one hop airborne network links), with the consideration of both historic and predicted states.

\begin{equation}
    F=\sum_{k=t-t'}^{t}N_G(k) + N_G(t+t')
    \label{Equation:Fitness}
\end{equation}
where $N_G$ is the total number of ground agents covered by an UAV and its neighbouring UAVs, calculated as below:
\begin{equation}
    N_G(t) = N(t) + \sum_{i=1}^{N_A}{|N_i(t)|}
\end{equation}

This fitness function considers the local coverage and the intermediate connected UAVs' coverage. It does not consider any indirectly connected UAVs' coverage, in order to save network bandwidth and computation cost. The GA aims to maximize the fitness; this means, to increase the local coverage as well as the number of neighbouring connections for establishing better connectivity.

\section{Experiments and Results} \label{Section:ExperimentResults}

\subsection{Experimental Setup}
The size of the environment is $1000 \times 1000$ units. In the environment operate 100 ground agents and 4 UAVs for a duration of 22000 time steps. The ground agents are initialized with random positions. The UAVs are initialized to form a $300\times300$ units square situated in the center of the environment, which means, a fully connected airborne network. There are a total of 30 random number generator seeds for initialing the ground agents in our experiments. Along with each ground agents initialization, each UAV initializes 50 chromosomes randomly (the size of the population in the GA) and then runs the GA repeatedly in $t'$ time windows. Therefore, there are total 30 runs for each scenario, to ensure statistical validity of the algorithm.

As mentioned earlier in the paper, two scenarios are used for the ground agents behavior to evaluate the proposed decentralized approach: classic boids and random walk. The parameter setting for each of the scenarios is listed in Table~\ref{tab:scenarios}.

\begin{table}[h]
    \caption{Two ground movement patterns, implemented via various force weights applied to ground agents.}
    \centering
    \begin{tabular}{|l|r|r|r|} \hline
        Scenarios & Cohesion & Alignment & Separation \\ \hline
        Classic Boids (CB) & 0.01 & 0.125 & 1 \\ \hline 
        Random Walk (RW) & N/A & N/A & 1  \\ \hline
    \end{tabular}
    \label{tab:scenarios}
\end{table}

The neighborhood for the ground agents is defined by $v_d=30$~units, and $v_\alpha=360$ degrees. The neigborhood for the UAVs is defined by the radius of coverage, which has the constant value $R=300$.

\subsection{Discussion of Results}

\begin{figure}
    \centering
    \includegraphics[width=0.35\textwidth]{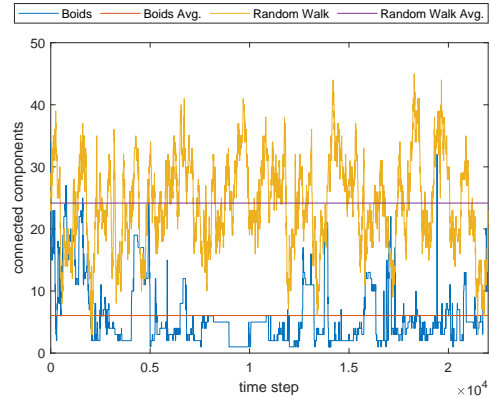}
    \caption{Connectivity comparison between boids and random walk}
    \label{Fig:com_con}
\end{figure}

The first set of results, illustrated in Figure~\ref{Fig:com_con}, shows the connectivity of the ground agents over time in the best cases, from both scenarios. It can be seen that, by using the proposed optimization algorithm, the swarm of UAVs is able to maintain high connectivity, i.e. there are very few disconnected network components. This results in very good survivability, with the number of connected components equal or close to equal to 1 most of the time in the boids case, and lower than 20 for the random walks case. Another aspect that can be observed in the boids case, is that from time to time there are spikes that show lower survivability. This is due to a bounce-back boundary condition, where agents reflect from the boundary of the environment. The bounce-back movement breaks the swarm behavior for a short period of time until the ground agents regroup in a new swarm formation. This shows that the proposed algorithm is able to recover from a large connectivity loss, and quickly provide high survivability for the newly formed swarm. The bounce-back movement has no visible effect in the case of random walks, since there is no movement pattern anyway.


\begin{figure}[h]
 \centering
    \begin{subfigure}[b]{0.35\textwidth}
        \includegraphics[width=\textwidth]{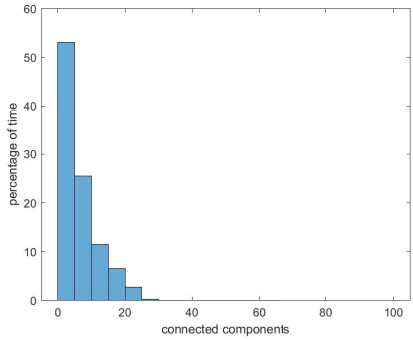}
        \caption{Boids}
        \label{fig:classic_confreq}
    \end{subfigure} \\
    \begin{subfigure}[b]{0.35\textwidth}
        \includegraphics[width=\textwidth]{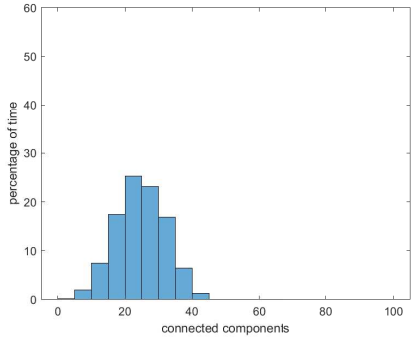}
        \caption{Random walk}
        \label{fig:random_confreq}
    \end{subfigure}
  \caption{The frequency of connect components from the best runs of boids and random walk}
  \label{Fig:ConFreq}
\end{figure}

Further, we are interested to see what is the amount of time a certain level of survivability is ensured, since the previous set of results did not show this very clearly. For the same simulation, i.e. the one with the best survivability results, we show in Figure~\ref{Fig:ConFreq} the percentage of time a certain connectivity was achieved. This is a remapping of the information presented in Figure~\ref{Fig:com_con}, to show a time summary of the connectivity measure. It can be seen that in the boids scenario the connectivity value is $1$ for more than $50\%$ of the time, which means the ground network is fully connected. For the random walk scenario, the graph shows lower, but still acceptable level of performance.

\begin{figure}[h]
 \centering
    \begin{subfigure}[b]{0.35\textwidth}
        \includegraphics[width=\textwidth]{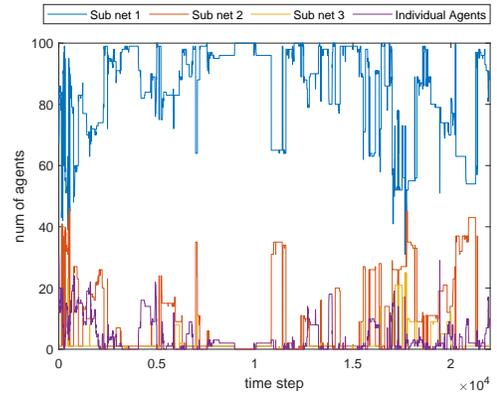}
        \caption{Boids}
        \label{fig:classic_line}
    \end{subfigure} \\
    \begin{subfigure}[b]{0.35\textwidth}
        \includegraphics[width=\textwidth]{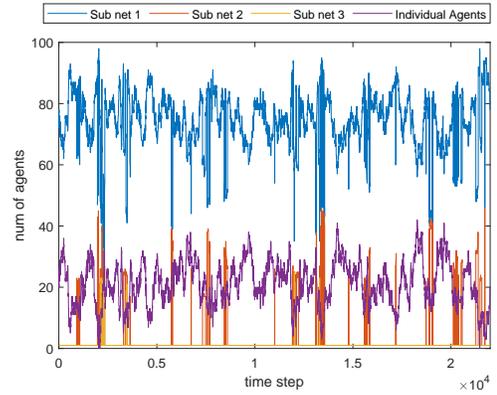}
        \caption{Random walk}
        \label{fig:random_line}
    \end{subfigure}
  \caption{The number of agents in TOP 3 largest sub-networks - from the best simulation run}
  \label{Fig:TOPNetsLines}
\end{figure}

The previous results showed the connectivity, which is measured as the number of connected components. However, they do not show the size of the connected components. Thus, for example, if a connectivity of 2 has been achieved, the two connected components may consist of 50-50 nodes, or they may consist of 95-5 nodes. The two examples are significantly different, with an obvious advantage of the latter case. Thus, the strength of the proposed approach may be overshadowed by this aspect, and the results shown in the previous figures may lead to the conclusion that the results are not a substantial achievement. We investigate this aspect by calculating the size of the largest three connected components, for the best runs. The results shown in Figure~\ref{Fig:TOPNetsLines} are convincing. It can be seen that the largest component contains over 70 nodes even in the random walk scenario, while in the boids scenario it contains most of the time over 90 nodes.


\begin{figure}[h]
 \centering
    \begin{subfigure}[b]{0.35\textwidth}
        \includegraphics[width=\textwidth]{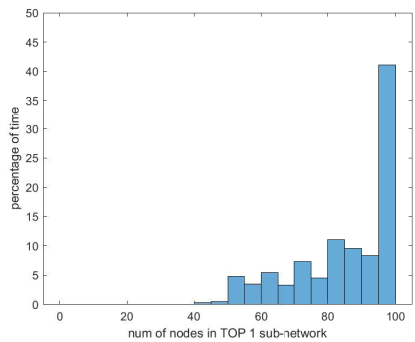}
        \caption{Boids}
        \label{fig:classic_top1}
    \end{subfigure} \\
    \begin{subfigure}[b]{0.35\textwidth}
        \includegraphics[width=\textwidth]{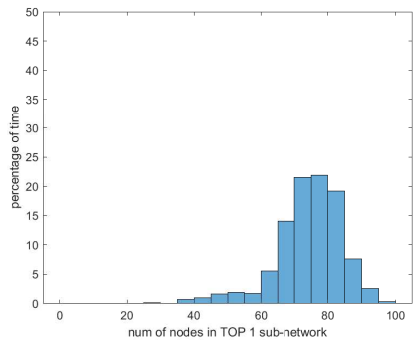}
        \caption{Random walk}
        \label{fig:random_top1}
    \end{subfigure}
  \caption{The number of agents over time (percentage) in the largest sub-network -  from the best simulation run}
  \label{Fig:TOPNetFreq}
\end{figure}

To further clarify what ``most of the time'' means, we show in Figure~\ref{fig:classic_top1} the amount of time (as percentage of the total simulation time) certain numbers of nodes are part of the largest connected component, in the best simulation run. This is in essence the same information presented in Figure~\ref{Fig:TOPNetFreq} for the largest component, remapped as a time summary. It can be seen, especially in the boids scenario, that over $40\%$ of the simulation time the largest component contains all the nodes, i.e. all the ground agents. However, the results are acceptable even in the random walk scenario, where the largest component contains over 70 nodes for more than $50\%$ of the time.

\section{Conclusions}\label{Section:Conclusions}

In this paper, we presented a nature-inspired approach to network survivability, which combines swarm intelligence and evolutionary computation. The proposed approach models the ground agents and the UAVs as a dual air-ground swarm that uses boids-like rules, and optimizes the movement of the UAVs using a decentralized real-time genetic algorithm. The proposed approach provides seamless integration of the ground and air swarms of agents, while also facilitating scalability and airborne responsivity to complex general ground behaviors.

The results obtained in simulations demonstrate that the mobility model used for the UAVs, and the associated evolutionary optimization algorithm are able to provide good levels of network survivability for complex ground movements, including the case of no movement pattern. 

We believe that the conceptual approach presented in this work can be successfully extended to a large variety of scenarios. In this paper we only used the three classic forces (cohesion, alignment and separation) that govern the emergence of swarming. However, numerous other behaviors (with their subsequent forces) can be investigated, such as leader following, flow field following, path following, obstacle avoidance, and many others. To the proposed approach more forces can be easily added, since the mobility model and the optimization algorithm depend mainly on the weights of these forces. Therefore, the approach is both size-wise and context-wise scalable. 

\bibliographystyle{IEEEtran}
\bibliography{CECbibfile}

\end{document}